# The Mikado Filesystem
## An experimental RPC filesystem running over gRPC

John D. Dougrez-Lewis
jlewis@lightblue.com
12th January 2024

## Abstract

Computer applications seeking to persist files remotely across the Internet today are faced with a bewildering choice of mechanisms which tend to boil down to monolithic proprietary closed-source Vendor solutions.

We introduce **The Mikado Filesystem** (*mikfs*), which provides an open simple lightweight interoperable portable extensible remote filesystem that is open source.

*mikfs* consists of client applications accessing remote servers via RPC running over TCP/IP network connections.

Rather than being described through a Protocol specification open to differing interpretations, *mikfs* is defined as a concrete set of API method calls over gRPC expressed in Google's Protocol Buffers' Interface Definition Language (IDL).

The gRPC infrastructure supports a wide variety of different programming languages and hardware platforms. For a given supported language + platform, the gRPC infrastructure is used to generate client-side and server-side stubs from the IDL which can then called from client and server code written in the selected client-side and server-side languages.

Thus, for instance, a client written in C# or java running on a Windows PC can access a server written in C++ running on a Linux box.

The Mikado Filesystem consists of a virtual hierarchical tree of files and directories. This logical filesystem is not constrained to the limits and file naming conventions of the server host's own physical native filesystem.

API methods are provided for authentication; for atomic file-level operations on files & directories; and for clients to register to receive notifications of file & directory changes on a server.

*mikfs* is extensible. Developers can write their own new server implementations and new client applications to communicate with existing clients and servers.

*mikfs* exposes a public API that any developer can provide an implementation for; allowing migration of hosted files between different implementations; allows for extension with new methods and features; is Open Source code available for inspection and adaptation;

gRPC provides secure authenticated connection & communication over HTTP/2, providing End-to-End Privacy & Security against eavesdropping of data in transit; provides Authenticated connections; *mikfs* provides support for multiple alternate user login mechanisms.

*mikfs* is provided as a source code distribution, '*The Bootstrap Distribution*', consisting of an ecosystem of clients, servers, tools and utilities. An initial set of Unit Tests is provided to assist with interoperation with existing clients and servers.

## Biographical Note

The author has been developing software since 1974 and is a mathematician by education. The choice of a default port of 9959 for *mikfs* is derived the author's birth date.

## Copyright & Licence

## Latest Version

The latest version of this paper can be found at:

https://bitbucket.org/jdougrezlewis/mikfs/src/master/documentation/mikfs%20paper.pdf

# 1. Introduction

Historically, and certainly up to the 1990s, computer applications persisted their documents on the local computer they were being run on.

The public popularity of Internet took off with the introduction of Tim Berners-Lee's World Wide Web [1].

With this rise of the Global Internet, information sharing across multiple disparate hosts has become the norm.

Whilst Internet file-sharing applications such FTP [2] existed prior to the Web, numerous technology stacks have been built up over HTTP and its successors to provide file-sharing solutions.

Despite this, people seeking to persist content remotely across the Internet today are faced with a bewildering choice of mechanisms which in the end typically boil down to monolithic proprietary closed-source Vendor solutions.

Application developers seeking to persist their content across the Internet face a number of issues:

If developers seek to base their remote persistence on monolithic proprietary closed-source Vendor solutions, then they are at the mercy of those Vendors: the choice of Vendor-controlled server venues Vendors offer to host, for a fee, or for free after entering into Mephistophelian bargains set out in their terms & conditions; and a lack of interoperability to migrate content to other Vendors.

If developers seek to base their remote persistence on Protocol specifications, then they need to code their clients to work with multiple different server implementations, each of which may differ in their interpretations of those specifications.

Solutions building over the top of vanilla HTTP introduce further complexity, e.g. file locking for read-write operations and user authentication. The risk of differences of interpretation increases with the complexity of the specification as further features are added on.

William of Ockham said "Entities must not be multiplied beyond necessity". Einstein said "Make everything as simple as possible, but not simpler". Saint-Exupéry said "Perfection is achieved, not when there is nothing more to add, but when there is nothing left to take away".

Publishing access to functionality as an API rather than a Protocol Specification reduces the scope for ambiguity by providing a deterministic concrete implementation that the developer has to integrate with.

It is far easier to code to invoke Unix/Linux System Calls [3] or a set of IBM PC BIOS interrupts [4] than it is to code a client to a protocol specification such as FTP.

Noting the above, we introduce **The Mikado Filesystem** (*mikfs*), which provides an open simple lightweight interoperable portable extensible remote filesystem that is open source.

*mikfs* consists of client applications accessing remote servers via RPC [5] running over TCP/IP network connections.

Rather than being described through a Protocol specification open to differing interpretations, *mikfs* is defined as a concrete set of API methods over gRPC [6][7] expressed in Google's Protocol Buffers [8][9] Interface Definition Language (IDL) [10].

The gRPC infrastructure supports a wide variety of different programming languages and hardware platforms.

For a given supported combination of language and platform, the gRPC infrastructure is used to generate client-side and server-side stubs from the IDL which can then be called from client and server code written in the selected client-side and server-side languages.

Thus a client written in C# or java running on a Windows PC can access a server written in C++ running on a Linux box.

The Mikado Filesystem consists of a virtual hierarchical tree of files and directories. This logical filesystem is not constrained to the limits and file naming conventions of the server host's own physical native filesystem.

API methods are provided for authentication; for atomic file-level operations on files & directories; and for clients to register to receive notifications of file & directory changes on a server.

*mikfs* is extensible. Developers can write their own new server implementations and new client applications to communicate with existing clients and servers.

*mikfs* exposes a public API that any developer can provide an implementation for; allowing migration of hosted files between different implementations; allows for extension with new methods and features; is Open Source code available for inspection and adaptation;

gRPC provides secure authenticated connection & communication over HTTP/2 [11][12][13], and provides End-to-End Privacy & Security against eavesdropping of data in transit;

*mikfs* provides extensible support for multiple alternate user login mechanisms.

*mikfs* is provided as a source code distribution, '*The Bootstrap Distribution*', consisting of an ecosystem of clients, servers, tools and utilities. An initial set of Unit Tests is provided to assist with interoperation with existing clients and servers.

# 2. Related Work

Remote Procedure Call (RPC) [5] has a long venerable history.

File Transfer Protocol (FTP) (1985) [2] was the mainstay on the Internet before the advent of HTTP.

Berners-Lee's original proposal (1989) [1] led the development of the WWW and HTTP.

Anderson (1996) wrote a very interesting theoretical paper on '*The Eternity Service*' [14].

HTTP Extensions for Distributed Authoring (WebDAV) appeared in RFC2518 (1999) [15], updated by RFC4918 (2007) [16].

A book (2003) [17] was written by the author of RFC4918. The latest WebDAV resources are can be found at [18].

Representational State Transfer (REST) appeared in Fielding's PhD dissertation (2000) [19].

The original WWW was conceived as Read-Write, but what grew out of it was a Read-Only web. Subsequent continuing research by Berners-Lee and his colleagues led to proposals for '*a Read-Write Web*' [20] and continue to this day with SOLID [21].

Various *commercial proprietary closed-source remote filesystems* have appeared: MediaFire [22], Storage Made Easy [23][24], Google [25], Microsoft [26], amongst others.

## 3. Contributions

This paper makes the following contributions:

1) We propose an Internet-wide RPC-based remote filesystem running over gRPC, exposed through an API, with the API expressed in language-independent IDL.

2) We implement the proposed filesystem [written in C#] as an ecosystem consisting of a cohesive suite of server and client applications and tools. The implementation is Open Source.

3) Our initial implementation serves as a harness to bootstrap further development & testing of future implementations and enhancements written in the wide variety of programming languages and platforms supported by gRPC and the IDL.

## 4. Background: History and Evolution

In 2013, the author of this paper wrote a web crawler application, Minos. In 2014, in order to organise & visualise the web page data harvested, a further application, Mikado, was developed.

Mikado provides a GUI to visualise graphs of links between web pages, to persist the graphs as "Mikado Documents"; to edit, update and create new documents, very much as one would handle word processor documents; and to view the web pages and document resources contained within those Mikado Documents.

Mikado could therefore said to be a "graph processor" application.

Mikado initially supported reading and writing of its documents to the local filesystem, and loading from documents hosted on the Internet on HTTP Web Servers.

While the local filesystem was Read-Write, the Internet was Read-Only. It soon became clear that what was required was to make the application Read-Write (RW) across the Internet.

Various file Read-Write mechanisms were considered, including FTP [2], WebDAV [15][16][17][18], and REST [19], but these were problematic because of difficulties finding suitable server implementations for various different platforms, inconsistencies arising from differences in server implementations, and having to implement client sides from Protocol Specifications from scratch.

Commercial Third Party vendor filesystems including MediaFire [22] and StorageMadeEasy [23][24] were considered, tried-out and in the end rejected; other monolithic solutions such as Google Drive [25] and Microsoft OneDrive [26] were rejected due to proprietary vendor lock-in concerns.

Apache Thrift [27] provided IDL-based RPC, but didn't provide the depth of bidirectional functionality required.

Then, in 2020, the author stumbled across gRPC [6] [7] which did provide the kind of bidirectional support required. Given the ecosystem of the wide variety of languages [28] and operating system platforms it supported, and being backed by the reputation of Google, the author decided to try it out.

The author created an API for a set of atomic file-level operations, expressed as RPC methods in Protocol Buffers IDL, and named it '*The Mikado Filesystem*' (or *mikfs* for short), and then coded an initial server application built using gRPC, and added support to the Mikado application as a *mikfs* client, so supporting the desired Read-Write Internet-enabled feature.

It soon became pretty clear that the *mikfs* functionality had outgrown its original parent application and has a much greater general utility & applicability, with *mikfs* being used as the basis of an API-based, language-neutral, platform-neutral, vendor-neutral Read-Write Internet filesystem.

So the author has coded an initial 'bootstrap' suite of server and client applications using the mikfs API, and that suite is hosted in this repo.

The intention of this 'bootstrap' suite is to allow and encourage other developers to embrace, enhance and extend new applications using the mikfs API across further languages and operating system platforms. The API itself is open to extension.

## 5. Design Principles

*mikfs* is a Remote Filesystem with the following features:

- A familiar Directory/Tree structure navigation

- Provides a deterministic API rather than a Protocol-based specification open to misinterpretation.

- A Virtual Filesystem not constrained to limits and file naming conventions of server's own host physical native filesystem.

- Atomic File-Level operations

- Notification of filesystem changes

- Privacy & Security of End-to-End Communication free from eavesdropping of data in transit

- Authenticated Access providing extensible support for alternate Authentication mechanisms.

- POSIX-like File Permissions for Authorisation

- Extended Attributes

- Support for Content-based Search Queries

- Support for Filesystem Replication between multiple alternate Hosting instances sharing the same API but different heterogeneous underlying implementations.

- Is not tied to a Proprietary Commercial Vendor.

- Is Open Source with code freely available for inspection and adaptation

- Has a public API expressed as IDL that is platform & programming language independent so that developers can implement servers & clients on different platforms & languages

- Has a public API that allows migration of hosted files between different implementations

- Has an extensible API to allow for new methods and features.

- Provides a self-contained ecosystem with a bootstrap implementation containing a set of Client & Server applications & utilities, together with Unit Tests that combined can demonstrate the system and bootstrap & ease development & validation of further alternative implementations against it.

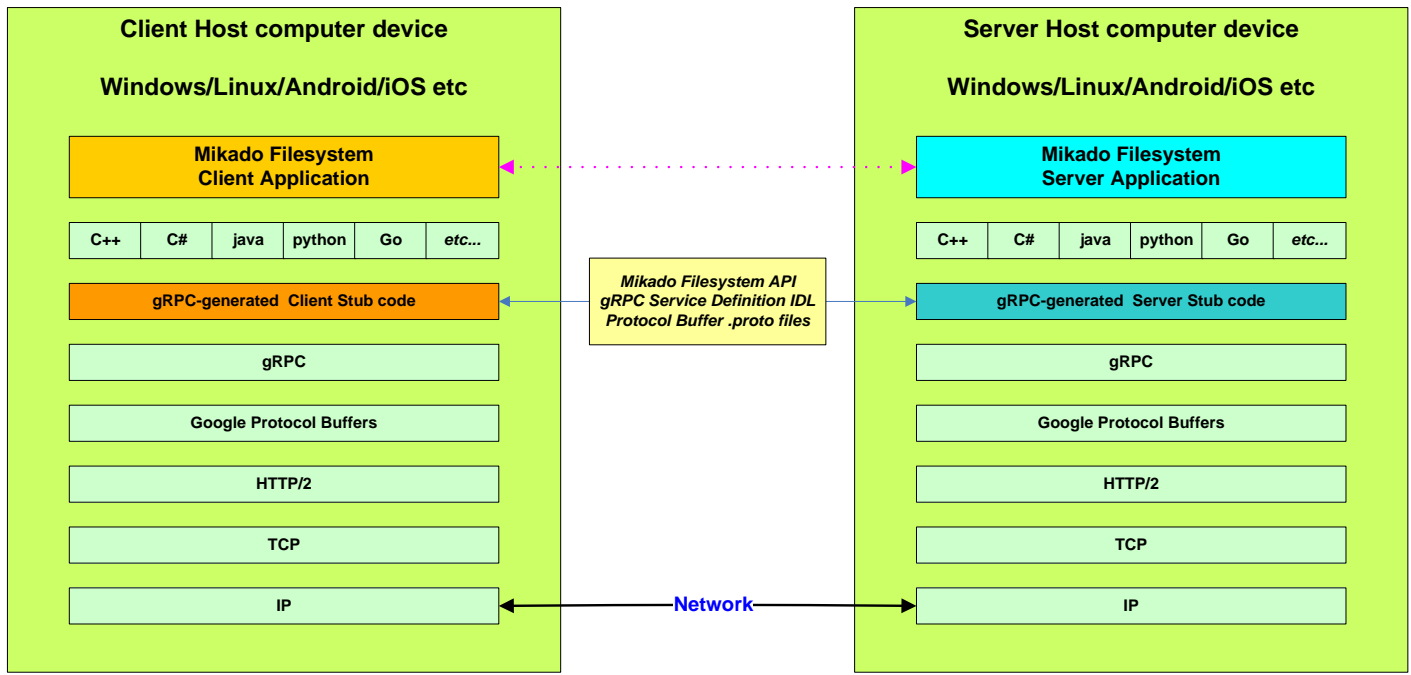

Figure: The mikfs gRPC Stack

## 6. API Methods

*mikfs* provides the following API Methods:

| #  | Method Name |
|----|-------------|
| 0  | GetApiInfo |
| 1  | Authenticate |
| 2  | Logout |
| 3  | GetHostWriteHandle |
| 4  | GetFile |
| 5  | GetFileInChunks |
| 6  | PutFile |
| 7  | PutFileInChunks |
| 8  | CreateDirectory |
| 9  | ReadDirectoryContents |
| 10 | MoveFile |
| 11 | CopyFile |
| 12 | MoveDirectory |
| 13 | CopyDirectory |
| 14 | DeleteFile |
| 15 | DeleteDirectory |
| 16 | GetDirectoryZip |
| 17 | GetDirectoryZipInChunks |
| 18 | CreateDirectoryUnzip |
| 19 | CreateDirectoryUnzipInChunks |
| 20 | SetPermissions |
| 21 | GetPermissions |
| 22 | UpdateAttributes |
| 23 | GetAttributes |
| 24 | FileSystemChangeSubscribe |
| 25 | Search |
| 26 | SearchSubscribe |

## 7. gRPC calling patterns utilised

gRPC methods support four calling patterns:

| |
|---|
| call with single request with single response |
| call with single request with stream of responses |
| call with stream of requests with single response |
| call with stream of requests with stream of responses |

mikfs uses:

- *streamed request streamed response* for the *Authenticate* method.
- *streamed request single response* for the *chunked* file write methods.
- *single request streamed response* for the *chunked* file read and subscribe methods.
- *single request single response* for the remaining methods.

## 8. A typical client-server session

A mikfs client-server session begins with the client application establishing a gRPC connection to the server.

The client then proceeds to call the *Authenticate* method initiating a Challenge-Response exchange with the server.

If the exchange completes successfully, the server provides a session handle to the client to be passed in future API calls.

The client can optionally call *GetApiInfo* to get information on the server, the particular API methods it supports, and other server information.

The client can optionally call *FileSystemChangeSubscribe* to be notified of file & directory change events on particular paths and filename patterns on the server.

In order to be able to make changes to the mikfs filesystem hosted on the server, the client calls *GetHostWriteHandle* to get a host write handle which can is combined with a client write handle to form an ownership handle to be passed across when writing files.

The client is then able to make API calls to conduct file write operations on the server.

Once the client has finished its file operations, it calls *Logout* to terminate the session, and then closes the gRPC connection.

Note that there is no concept of a 'current directory' on the server side during a session. This feature can be implemented by a client application if it requires one.

## 9. Authentication

The mikfs client-server session begins with the client application establishing a gRPC connection to the server.

In gRPC, communication between server and client is over HTTP/2 [11][12][13] which using TLS [29][30][31] to establish and provide secure reliable end-to-end encrypted transport of data, preserving its integrity.

Once a gRPC connection has been established, the client calls the *Authenticate* method to initiate a Challenge-Response exchange with server.

The contents of the Challenge-Response exchange, passed in the clear at the API call level, are secure from eavesdropping when passed over the wire in the TLS-protected gRPC connection.

The message exchange can support an extensible set of multiple Challenge-Response schemes. The server selects a particular scheme to use and specifies its chosen scheme at the start of the exchange. The client then has to respond to the scheme chosen.

The bootstrap implementation supports two 'well-known' schemes: '*Durin*' and '*UserPassword*'. See the diagram below for an exchange of messages under these schemes.

## 10. X.509 Certificates

TLS can use X.509 Certificates [32] to validate the identities of the server and client

The gRPC implementations for various different language & platforms provide means of specifying the X.509 Certificates when creating & establishing a connection between a server and a client.

The bootstrap mikfs server implementation and client applications allow users to specify the X.509 Certificate details for the server, and optionally, for clients.

X.509 Certificates can be created & signed by an external Certificate Authority (CA) or created and self-signed by users.

Lets Encrypt [33][34] is a CA that allows developers to create X.509 Certificates that are signed by Lets Encrypt for no charge. The Certificates are valid for 90 days.

Other commercial Certificate Authorities exist who will create Certificates with a longer validity for their customers for a fee.

The Mikado Filesystem Installation binary installs a standalone UI application to create and configure self-signed certificates for use with locally hosted servers & clients.

## 11. Directory Zips

To ease replication of directory trees, the Mikado Filesystem provides methods to Zip and Unzip *mikfs*-hosted directories.

## 12. Filesystem instance mutability

*mikfs* server filesystem instances can be started as Read-Write, Read-Only or Append-Only.

## 13. The Ownership Handle Concept

*mikfs* uses the concept of Ownership Handles to control access to file content hosted on remote mikfs servers.

Unlike other server access systems, the control of Ownership Handles is not centralised on the server instances themselves, but is, instead, split between the clients and the servers.

An Ownership Handle consists of two parts – a Group Handle allocated & controlled by the server host and a Group Handle allocated & controlled by the client that originated the file content.

When a client creates new content in the form of a file or directory, to compose an Ownership Handle, they must allocate a client (user) Group Handle and combine it with a server (host) Group Handle previously retrieved from the server in the current session using the *GetHostWriteHandle* API call. The Ownership Handle is then stored on the server as an attribute of the file or directory.

The Ownership Handles stored on the server are then private to the server and are not otherwise retrievable from it.

To determine a caller's access permissions for an existing file or directory on a remote server, when calling the mikfs API method, the client must present the server with an Ownership Handle.

Because the client needs knowledge of Ownership Handles, these are required to be cached on the client side and mechanisms provided to store and retrieve them, and to distribute & share them between clients who are to access the content.

The clients include the creating process, other instances of the process and other client applications on the same machine or other machines elsewhere on the Internet.

To accommodate caching of, and access to, Ownership Handles on the local network, the mikfs bootstrap implementation provides three ImportExport metadata servers - one command-line based, a GUI-based server, and one that runs as a Windows Service. An ImportExport server is started automatically (if not already running) when a bootstrap client is started.

Client applications query and update the Ownership Handles held on the ImportExport servers via a gRPC API. This then allows access to Ownership Handles for both local and remote clients.

This provides an abstraction for synchronised access to the cache of Ownership Handles, and the cache itself then becomes an implementation detail.

Typically the remote clients would be expected to be restricted to the user's local network rather than the whole Internet, creating a requirement to establish and ensure firewall controls to prevent unwanted external access.

Where content is intended to be made and distributed for general read availability, the requirement for exact matching Ownership Handles can be avoided by setting Read Group Permissions on the content to allow access to other users outside of the exact matching User Group and Host Group.

Alternatively, via the Mikado Filesystem Explorer application, content-creating users can Export the ImportExport Ownership Handle metadata to a file, send or publish the file, making it available to other users who can then use the Mikado Filesystem Explorer to import the metadata to their own local ImportExport server, and then access the content.

The Mikado Filesystem Explorer Import/Export mechanism allows filtering of files and restrictions masking the User and/or Host Group Handle parts of Ownership Handle metadata exported.

## 14. File Permissions

*mikfs* uses a file permissioning scheme similar to POSIX file permissions [35], but extends the bitmask to include both server-side-supplied and client-side-supplied components embedded in an **Ownership** handle.

**Ownership structure**

| Field Name | Description | Data representation |
|---|---|---|
| HostGroup | Represents server-supplied group handle | GroupOwner structure |
| UserGroup | Represents client-supplied group handle | GroupOwner structure |

**GroupOwner structure**

| Field Name | Description | Data representation |
|---|---|---|
| Key | Ownership Identifier | Array of bytes |

**Determining Permission Group**

To determine a client's access rights to a file or directory used in an API call, a client user specifies an Ownership handle the API call. The client ownership is then compared with the file/directory node's stored Ownership handle and one of the 4 following group memberships is then determined.

| Membership | Description |
|---|---|
| Owner | Caller's HostGroup matches node's HostGroup, and Caller's UserGroup matches node's UserGroup |
| UserGroup | Caller's HostGroup doesn't match node's HostGroup, but Caller's UserGroup matches node's UserGroup |
| HostGroup | Caller's HostGroup matches node's HostGroup, but Caller's UserGroup doesn't match node's UserGroup |
| Other | Caller's HostGroup doesn't match node's HostGroup, Caller's UserGroup doesn't match node's UserGroup |

For a match to occur, either one or both of the Caller's and Node's Key must be of zero length (a "Matches All" wildcard match), or else the two keys must be of equal non-zero length and match exactly byte-for-byte.

File/Directory node permissions are then analogous to the POSIX file permissions *owner*-*group*-*world*, but instead using the above four names, namely *owner*-**usergroup**-**hostgroup**-*other*.

Each group then has bit permissions for read-write-execute.

The Mikado Filesystem has no concept of an executable file – all files are considered to be data, so the "execute" flag is unused for files, and only used for directories.

In addition, there is a "Sticky" bit, analogous to POSIX, for directories, used to control owner-only deletion of files in directories.

## 15. Attributes

The Mikado Filesystem provides a system of attributes and custom (extended) Attributes for files & directories.

Each file and directory node in the Mikado file system has the following standard attributes:

| Attribute | Description |
|---|---|
| Size | Size of File in bytes (0 for directories) |
| LastModifiedTime | nanoseconds since 01/01/1970 00:00:00 UTC |
| PermissionsMask | POSIX-like File/Directory access flags |
| Owner | Handle identifying ownership |
| CustomAttributes | Collection custom extended attributes. |

*Size* & *LastModifiedTime* are set on API file write operations, and retrievable via the API method *ReadDirectoryContents*.

*PermissionsMask* is set in file/directory creation operations and *SetPermissions*, and retrievable via *ReadDirectoryContents* and *GetPermissions*.

*Owner* handles for files are held the client side and passed & set in file/directory creation operations.

New *Owner* handles are constructed with the help of the user calling *GetHostWriteHandle*.

*Owner* handles are used on the server side to determine whether clients calling the API are permissioned for a file/dir operation.

*Owner* handles, once set, are therefore sensitive for security & cannot be retrieved by a client of the API.

The client side of the API is required to maintain a persisted cache of the *Owner* handles for the files they store on the server. This is supported by implementing higher-level abstractions in a mikfs client implementation.

*Custom Attributes* are set with *UpdateAttributes* and can be retrieved with *GetAttributes*.

A Custom Attribute consists of a user-definable name and value. Names are not case-sensitive and should be canonicalised to Lower Case on the server side, and returned in their canonicalised form.

A file/directory's attributes can only be created, set & updated by its Owner.

## 16. Virtual Filesystem

*mikfs* servers implement a Virtual Filesystem unconstrained by the limits and file naming conventions of the host server's own physical native filesystem.

| Parameter | Value |
|---|---|
| Valid File name characters | Any Unicode char except '\0' and '/' |
| Directory separator character | '/' |
| Min File/Directory name length | 1 character |
| Max File/Directory name length | 255 characters |
| Max Full Path length | 4095 characters |
| Min File size | 0 bytes |
| Max File size | $2^{31}-1$ bytes |
| Extended Attribute name length | 1 to 255 characters |
| Extended Attribute value length | 0 to $2^{16}-1$ (65,535) characters |

## 17. Further Notes on Security

Whilst the use of TLS in the gRPC HTTP/2 layer prevents eavesdropping of the contents of data sent across the wire, this does not prevent suitably equipped external third parties from identifying the creation of socket connections between the client and server, or observing the quantity of encrypted data travelling across the connections.

Further, the bootstrap implementation does not encrypt data at rest on the server.

There is scope in future to encrypt data at rest held on the server, but for this to be secure, the encryption/decryption keys must not be stored on server and should instead be passed transiently from the client within API method arguments.

Alternatively, such functionality, if required, could be provided externally at the application level, e.g. through the use of one-time PADs [36] for paired clients and servers.

## 18. ImportExport service

To support caching of, and access to, Ownership Handles on the local network, the mikfs bootstrap implementation provides two ImportExport metadata servers running as gRPC services.

The ImportExport service provides the following API Methods:

| #  | Method Name          |
|----|----------------------|
| 0  | AddSite              |
| 1  | AddSites             |
| 2  | AddPath              |
| 3  | AddPaths             |
| 4  | GetSites             |
| 5  | GetPath              |
| 6  | GetPathsForSite      |
| 7  | GetPathsForAllSites  |
| 8  | RemoveSite           |
| 9  | RemoveSites          |
| 10 | RemoveAllSites       |
| 11 | RemovePath           |
| 12 | RemovePaths          |
| 13 | SitesSubscribe       |

## 19. Search service

To support the discovery across the Internet of mikfs server instances and document content hosted on those servers, the bootstrap release provides some experimental work-in-progress Search Engine functionality implemented in the Mikado Search Engine Server application and incorporated into the client-side applications included in the release.

The Mikado Search Engine Server runs as a gRPC service.

The service provides the following API Methods:

| # | Method Name                          |
|---|--------------------------------------|
| 0 | GetApiInfo                           |
| 1 | SearchForDocumentRecords             |
| 2 | SubscribeSearchForDocumentRecords    |
| 3 | InsertDocumentRecords                |

## 20. Default IP Port Numbers

The following IP Port numbers are used by the gRPC services in the Mikado ecosystem:

| Service              | Port |
|----------------------|------|
| *mikfs*              | 9959 |
| Mikado Search Engine | 9960 |
| ImportExport Service | 9961 |

If it is intended to use these services and expose then across to the Internet, then these are the Port Numbers people might reasonably expect the servers to be found at. These port numbers are not mandatory.

If it is intended these run these services privately (only to be used and exposed on a local intranet or a single machine), then other non-guessable Port Numbers ought to be selected, and suitable Firewalls configured to prevent access from the Public Internet.

In particular, it is highly unlikely that an ImportExport Service would ever want to be accessible from the Public Internet, or even from other machines on the same Intranet.

The Configuration Creator UI application distributed with the Mikado Bootstrap Install to create a working configuration sets up randomised Port Numbers.

## 21. Applications in the Bootstrap Release

The current mikfs 'bootstrap' implementation is written in C++ and C# and contains a suite of mikfs server & client applications which can be built to run on a Windows PC running Windows 7 or later.

The 'bootstrap' applications access the *mikfs* API via a C# stub layer to gRPC.

The following applications are provided:

| Binary Name                            | Description                                                                                                                 |
|----------------------------------------|-----------------------------------------------------------------------------------------------------------------------------|
| MikadoFileExplorer.exe                 | UI-based Mikado Filesystem Explorer client and Local Mikado Filesystem server.                                              |
| MikFSServer.exe                        | Command line console-based Mikado Filesystem server.                                                                        |
| Mikado.exe                             | UI-based Graph Document Editing application using Mikado Filesystems for remote storage and Mikado Search Engines for queries. |
| MikFSConfigCreatorUI.exe               | UI-based Utility for automated creation & configuration of User-Password and self-signed X.509 certificates after suite installed. |
| MikadoPerf.exe                         | UI-based Performance measuring tool.                                                                                        |
| mush.exe                               | The Mundane Shell. Command line console-based mikfs client.                                                                 |
| MikadoSearchClientUIApp.exe            | UI-based Standalone Client for querying Mikado Search Engine servers.                                                       |
| MikadoSearchEngineServer.exe           | Command line console-based Mikado Search Engine server.                                                                     |
| *MikFSImportExportServerGUI.exe*       | *UI-based permissions metadata server.*                                                                                     |
| *MikFSImportExportServer.exe*          | *Command line console-based permissions metadata server.*                                                                   |
| *MikFSServer WindowsService.exe*       | *Windows Service-based Mikado Filesystem server.*                                                                           |
| *MikadoSearchEngine WindowsService exe* | *Windows Service-based Mikado Search Engine server.*                                                                        |
| *MikFSImportExport WindowsService.exe* | *Windows Service-based permissions metadata server.*                                                                        |
| testhost.exe                           | *Client-side Unit Tests.*                                                                                                   |

## 22. *mikfs* URLs

The bootstrap applications allow access to ***mikfs***-hosted files via URLs of the standard form:

```
mikfs://host:port/path/filename
```

Note however that spaces and other pathological characters are permitted in mikfs paths & filenames, so care should be taken when parsing such URLs, and passing them on command lines.

## 23. The *mikfs* C++ stack

The API for the Mikado Filesystem is defined by the Protocol Buffers *.proto* IDL files for the *mikfs* service.

The IDL files allow for *mikfs* clients and servers to be built in any language and run on any platform supported by the gRPC tools.

The Mikado Filesystem Bootstrap suite also includes support for a *mikfs* C++ code stack that runs on both Windows and Linux.

Building the *mikfs* C++ code stack generates code stubs for both clients and servers, allowing C++ clients and servers to be built and run on Windows and Linux, to interact with other Bootstrap suite & user-developed servers and clients.

The *mush_cpp* client is a C++ port of the C# *mush* application ("*The Mundane Shell*"), an interactive ftp-like command-line console application for transferring files between mikfs servers and clients.

The *mush_cpp* source code demonstrates how to access the *mikfs* API in C++, and serves as a starting for users seeking to develop their own C++ *mikfs* client applications.

No C++ server application is currently provided, but the generated C++ code stubs provide the necessary building blocks for starting on this, and the C# server implementation is ripe for porting.

## 24. Conclusion and Future Work

We have presented an Internet-accessible filesystem called The Mikado Filesystem (*mikfs*) which runs over gRPC.

We have provided the bootstrap release:

- to provide a viable functional & documented ecosystem:

- to allow for the development, testing & bug fixing of that release;

- to receive developer and user feedback for the current implementation;

- to allow others to participate in the development and testing further server implantations and client applications across diverse languages, hardware and operating systems that are currently supported by the gRPC stack;

- to allow others to participate in the evolution, extension and refinement of the Mikado Filesystem;

- to allow users to exercise and become familiar with the system;

- to generate feedback from users and developers for extending to add new features;

*The Bootstrap Distribution* is an experimental continuing work-in-progress.

## 25. Further Details

The *mikfs* source repository is located at:

https://bitbucket.org/jdougrezlewis/mikfs

Further documents are located in the repo's ./documentation subdirectory:

The latest version of this paper: mikfs paper

Installing the Mikado Filesystem

Building the Mikado Filesystem

Building the mikfs C++ stack and the mush_cpp client

Mikado Filesystem Explorer User Guide

The Mikado Filesystem Distilled

The Complete and Utter Mikado Filesystem

The Mikado User Guide

Things to Come

# Appendix: Authentication Flow examples

The *mikfs Authenticate* API call consists of an exchange of a sequence of Challenge-Response messages between the client and server to try to establish an authenticated session. The Bootstrap release implementation of the Mikado Filesystem currently supports two authentication schemes, "*Durin*" and "*UserPassword*".

The sequence is initiated by the client and continues until the server grants a session handle or returns a failure status.

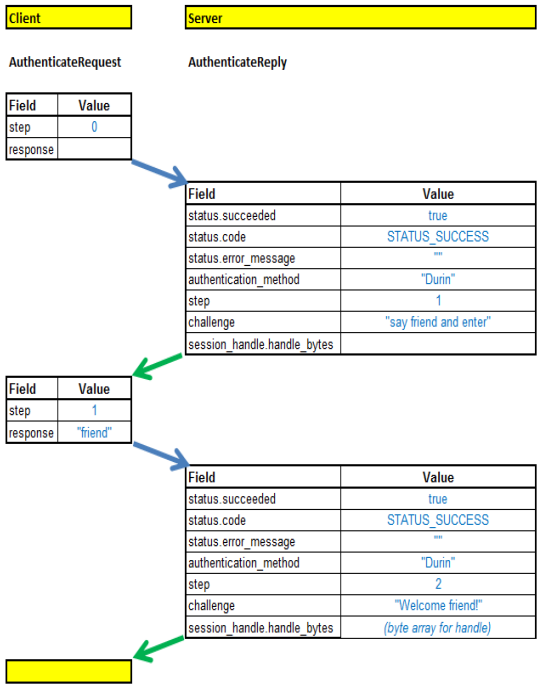

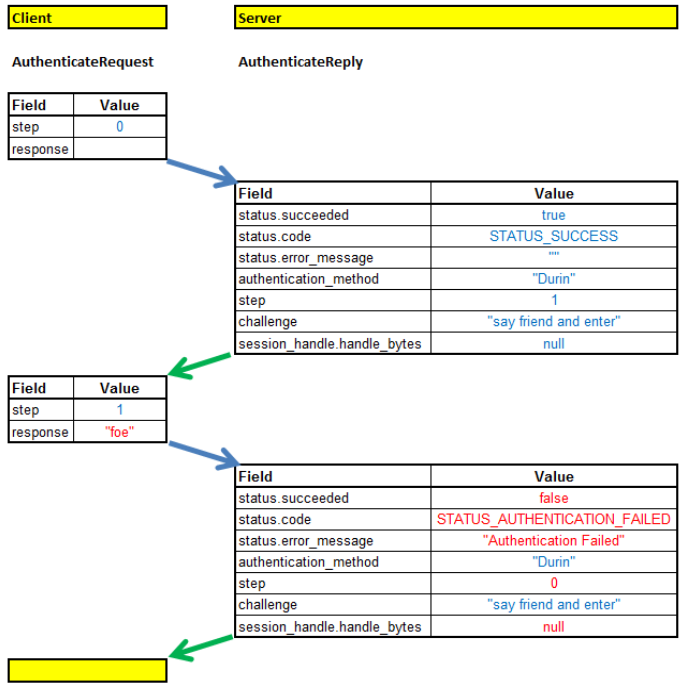

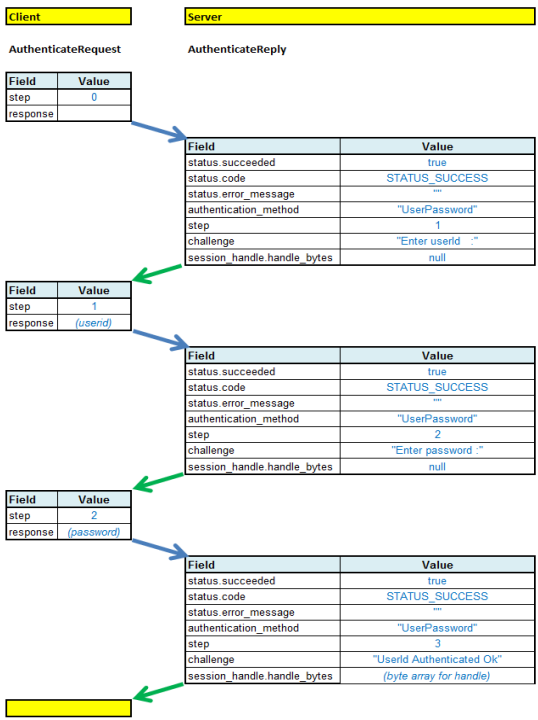

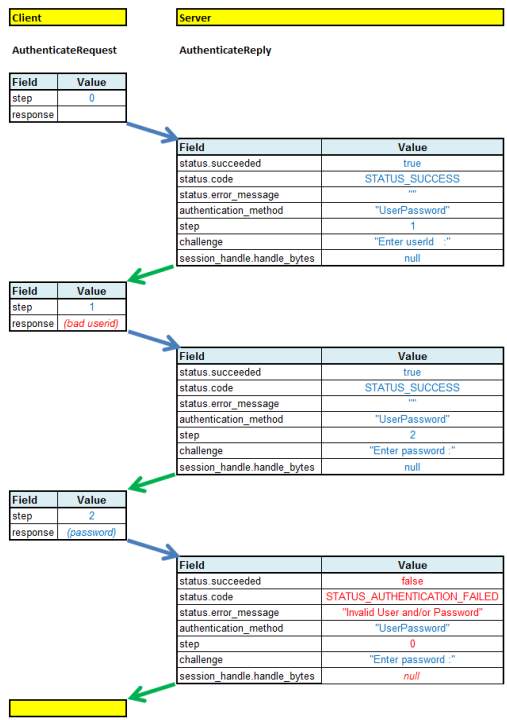